\documentclass[11pt,a4paper]{article}
\pdfoutput=1

\usepackage{jheppub}
\usepackage{latexsym}
\usepackage{revsymb}
\usepackage{multirow}
\usepackage{color}
\usepackage[usenames,dvipsnames,svgnames,table]{xcolor}

\usepackage{graphicx}
\usepackage{epsfig}  
\usepackage{epsf}    
\usepackage{dcolumn}
\usepackage{bm}
\usepackage{dcolumn}
\usepackage{textcomp}
\usepackage{float}
\usepackage{hypcap}
\usepackage[]{hyperref}
\usepackage{adjustbox}
\usepackage{multirow}

\usepackage{amsmath}
\usepackage{subfigure}
\usepackage{alphalph}
\usepackage{morefloats}
\usepackage{textcomp}
\usepackage[utf8]{inputenc}

\newcommand{\be}{\begin{equation}}
\newcommand{\ee}{\end{equation}}
\newcommand{\ba}{\begin{eqnarray}}
\newcommand{\ea}{\end{eqnarray}}

\newcommand{\cnv}{Cherenkov}

\preprint{FTUAM-19-25 \vspace{-3.5mm} \begin{flushright} IFT-UAM/CSIC-19-171 \end{flushright}}

\title{Sensitivity to light sterile neutrinos at ESSnuSB} 

\author[a,b]{Monojit Ghosh,}
\author[a,b,c]{Tommy Ohlsson,}
\author[d]{Salvador Rosauro-Alcaraz}

\affiliation[a]{Department of Physics, School of Engineering Sciences, KTH Royal Institute of Technology,\\ AlbaNova University Center, Roslagstullsbacken 21, SE--106 91 Stockholm, Sweden }
\affiliation[b]{The Oskar Klein Centre, AlbaNova University Center, Roslagstullsbacken 21,\\ SE--106 91 Stockholm, Sweden}
\affiliation[c]{University of Iceland, Science Institute, Dunhaga 3, IS--107 Reykjavik, Iceland}
\affiliation[d]{Departamento de F{\'i}sica Te{\'o}rica and Instituto de F{\'i}sica Te{\'o}rica, IFT-UAM/CSIC,\\ Universidad Aut{\'o}noma de Madrid, Cantoblanco, 28049, Madrid, Spain}

\emailAdd{manojit@kth.se}
\emailAdd{tohlsson@kth.se}
\emailAdd{salvador.rosauro@uam.es}

\abstract{
We present a comprehensive analysis in the 3+1 active-sterile neutrino oscillation scenario for the sensitivity of the ESSnuSB experiment in the presence of light sterile neutrinos assuming both a far (FD) and a near (ND) detector. Our analysis show that when the ND is included, the results are significantly different compared to the ones obtained with the FD only. We find that the capability of ESSnuSB to constrain the sterile mixing parameters is $\sin^22\theta_{\mu e} \sim 10^{-4}$ for $\Delta m^2 = 1$ eV$^2$ if the ND is included and it becomes $\sin^22\theta_{\mu e} \sim 10^{-2}$ without the ND. Furthermore, we show that the sensitivity can go down to $\sin^22\theta_{\mu e} \sim 10^{-3}$ for the most conservative choice of the systematics on the ND. Comparing the sensitivity with T2HK, T2HKK, and DUNE by considering the FD only, we find that the sensitivity of ESSnuSB is smaller for most of the parameter space. Studying the CP violation sensitivity, we find that if the ND is included, it can be larger in the 3+1 scenario than in the standard one. However, if the ND is not included, the sensitivity is smaller compared to the one in the standard scenario. We also find that the CP violation sensitivity due to $\delta_{13}$ is larger compared to the one induced by $\delta_{24}$. The sensitivities are slightly better for the dominant neutrino running ratio of ESSnuSB. 
}

\arxivnumber{}
\begin{document}
\maketitle
 
\section{Introduction}
\label{sec1}

In the standard three-flavor scenario, the phenomenon of neutrino oscillations is described by six parameters. There are three leptonic mixing angles $\theta_{12}$, $\theta_{23}$, and $\theta_{13}$, two mass-squared differences $\Delta m^2_{21}$ and $\Delta m^2_{31}$, and one Dirac CP-violating phase $\delta_{13}$. Among the six parameters, $\theta_{12}$, $\theta_{13}$, $\Delta m^2_{21}$, and $|\Delta m^2_{31}|$ are measured with very good precision \cite{Esteban:2018azc}. At present, the unknown parameters are: (i) The neutrino mass ordering or the sign of $\Delta m^2_{31}$, which can be either $\Delta m^2_{31} > 0 $ giving normal mass ordering (NO) or $\Delta m^2_{31} < 0 $ giving inverted mass ordering (IO). (ii) The value of $\theta_{23}$ and to which octant it belongs, which can be either  $\theta_{23} < 45^\circ$ in the lower octant (LO), $\theta_{23} > 45^\circ$ lying in the higher octant (HO), or $\theta_{23} = 45^{\circ}$ resulting in maximal mixing. (iii) The value of the CP-violating phase $\delta_{13}$. In a quest to determine these unknown parameters, the currently running experiments are T2K \cite{Abe:2019vii} and NO$\nu$A \cite{Acero:2019ksn}. The data of these experiments suggest that there is a mild tendency towards NO, HO of $\theta_{23}$, and $\delta_{13}$ around $-90^\circ$. There are many neutrino oscillation experiments, which are proposed to measure these unknown parameters at very significant confidence level.   
 
Apart from the standard three-flavor neutrino oscillation scenario, it is also possible to probe various new physics scenarios in neutrino oscillation experiments. An example of one such scenario is the existence of a light sterile neutrino. Sterile neutrinos are SU(2) singlets which do not interact with the Standard Model particles. However, they can mix with the active neutrinos and therefore take part in the oscillation phenomenon. Recently, there are compelling experimental evidences towards the existence of a light sterile neutrino at the eV scale \cite{Boser:2019rta}. For a review on this subject, we refer to Ref.~\cite{Abazajian:2012ys}. Probing light sterile neutrinos in long-baseline neutrino oscillation experiments is a very interesting topic. In such experiments, protons are collided on a fix target to produce pions and then the pions decay to produce muon and antimuon neutrinos having energies from hundreds of MeV to a few GeV. These neutrinos are detected by a far detector (FD) at a distance which is suitable to study neutrino oscillations that are governed by $\Delta m^2_{31}$. At this distance, the oscillations driven by sterile neutrinos, which would be governed by a neutrino mass-squared difference around 1~eV$^2$ and therefore averaged out, can affect the physics measurement at the FD. Apart from that if there is a near detector (ND), then it is also possible to probe the oscillations that are governed by the light sterile neutrino mass-squared differences. Analyses of sterile neutrinos in the context of long-baseline neutrino oscillation experiments have been carried out previously in the literature \cite{Berryman:2015nua, Gandhi:2015xza, Agarwalla:2016xxa, Agarwalla:2016xlg, Coloma:2017ptb, Choubey:2017cba, Choubey:2017ppj, Haba:2018klh, Ghosh:2017atj, Agarwalla:2019zex, Majhi:2019hdj, Reyimuaji:2019wbn, Dentler:2018sju}.

In this work, we study the sensitivity to a light sterile neutrino in the context of the ESSnuSB experiment \cite{Baussan:2013zcy,Wildner:2015yaa}. ESSnuSB is a proposed long-baseline neutrino oscillation experiment in Sweden. In this experiment, the neutrinos will be produced by a proton beam of energy 2.5~GeV. This experiment is mainly designed to measure $\delta_{13}$ at the so-called second oscillation maximum. We address two important questions: (a) how well ESSnuSB can put bounds on the sterile mixing parameters assuming there are no sterile neutrinos in the data and (b) how the CP measurement capability of ESSnuSB is affected assuming the existence of light sterile neutrinos in the data. We perform our study by using three detector configurations of ESSnuSB: (i) combined FD and ND with correlated systematics, (ii) combined FD and ND without any spectral information on the ND, and (iii) an FD only with an overall systematic uncertainty. Here we would like to mention that Ref.~\cite{Agarwalla:2019zex} has discussed (b) assuming an FD only. However, as we mentioned earlier, with an FD only, the oscillations of sterile neutrinos with $\Delta m_{41}^2\sim 1$~eV$^2$ are averaged out. Only in the presence of an ND, one can probe the oscillations of sterile neutrinos. Therefore, the results in the presence of both an FD and an ND can be different from the results obtained with an FD only. 

This work is organized as follows. In Section~\ref{sec2}, we will discuss neutrino oscillations in the 3+1 active-sterile scenario. Then, in Section~\ref{sec3}, we will give the configuration of ESSnuSB, which is used in our analysis. Next, in Section~\ref{sec4}, we will present our results, and finally in Section~\ref{sec5}, we will summarize and conclude.

\section{The 3+1 active-sterile neutrino oscillation scenario}
\label{sec2}

In general, in the 3+1 active-sterile neutrino oscillation scenario, the neutrino oscillation probability from a neutrino flavor $\nu_{\alpha}$ to another one $\nu_{\beta}$ is given by (see e.g.~Ref.~\cite{Giunti:2007ry})
\begin{equation}
P_{\alpha \beta} \equiv P(\nu_\alpha \to \nu_\beta) = \sum_{i, j=1}^{4}U^*_{\alpha i}U_{\beta i}U_{\alpha j}U^*_{\beta j} \exp\left(-{\rm i} \frac{\Delta m_{i j}^2 L}{2E}\right)\,, \quad \alpha,\beta = e,\mu,\tau,s\,,
\label{eq:probab}
\end{equation}
where $\Delta m_{i j}^2 \equiv m_i^2-m_j^2$ is the definition of the neutrino mass-squared differences, $m_i$ being the mass of the neutrino mass eigenstate $\nu_i$ ($i = 1,2,3,4$) as well as $L$ and $E$ are the baseline length and the neutrino energy, respectively. Furthermore, the quantities $U_{\alpha i}$ are the leptonic mixing matrix elements. Thus, the full $4 \times 4$ leptonic mixing matrix $U$ can be parametrised through successive two-dimensional rotations as \cite{Gandhi:2015xza}
\begin{align}
U =  U_{34}(\theta_{34},\delta_{34})
U_{24}(\theta_{24},\delta_{24})
U_{14}(\theta_{14},0)
U^{3\nu}\,, \\
U^{3\nu}
=
U_{23}(\theta_{23},0)
U_{13}(\theta_{13},\delta_{13})
U_{12}(\theta_{12},0)\,,
\end{align}
where $U_{ij}(\theta_{ij},\delta_{ij})$ denotes a rotation in the $(i,j)$-plane with mixing angle $\theta_{ij}$ and CP-violating phase $\delta_{ij}$ and $U^{3\nu}$ corresponds to the ordinary three-flavor leptonic mixing matrix. In order to simplify the notation, we use the two following abbreviations
\begin{equation}
\mathrm{c}_{ij} \equiv \cos\theta_{ij} \quad \text{and} \quad \mathrm{s}_{ij}=\sin\theta_{ij}\,.
\end{equation}
For small values of $L/E$, i.e.~relevant at the ND, assuming $\Delta m_{4i}^2 \gg \Delta m_{31}^2,\Delta m_{21}^2$ ($i=1,2,3$), the neutrino oscillation probability from Eq.~(\ref{eq:probab}) is dominated by the effect of sterile neutrinos, since ordinary three-flavor neutrino oscillations have not yet developed. Therefore, an effective two-flavor approximation can be used (see e.g.~Ref.~\cite{Boser:2019rta}) to describe it as
\begin{equation}
P^{\text{ND}}_{\alpha \beta} \simeq \delta_{\alpha \beta}  + (-1)^{\delta_{\alpha \beta}}\sin^22\theta_{\alpha \beta} \sin^2\frac{\Delta m^2_{41} L}{4 E}
\label{Eq:Prob_SBL}
\end{equation}
with the effective amplitude
\begin{equation}
\sin^22\theta_{\alpha \beta} \equiv 4 |U_{\alpha 4}|^2 \left|\delta_{\alpha \beta} - |U_{\beta 4}|^2\right|\,.
\label{eq:ampl}
\end{equation}
The approximation in Eq.~(\ref{Eq:Prob_SBL}) is very useful to help understand most of the phenomenology of oscillations with sterile neutrinos, except for CP violation which cannot be described using a two-flavor approximation. In super beam neutrino experiments neutrino oscillations are probed through the \emph{appearance} ($\nu_\mu \rightarrow \nu_e$) and \emph{disappearance} ($\nu_\mu \rightarrow \nu_\mu$) channels.
Therefore, using Eq.~(\ref{eq:ampl}), the mixing parameters that affect these oscillations are the amplitudes $\sin^22\theta_{\mu e}$ and $\sin^22\theta_{\mu \mu}$, which are given by 
\begin{align}
\sin^22\theta_{\mu e} &= 4 |U_{e4}|^2 |U_{\mu 4}|^2 = \mathrm{s}^2_{24}\sin^2 2\theta_{14}\,, \\
\sin^22\theta_{\mu \mu} &= 4 |U_{\mu 4}|^2 (1 - |U_{\mu 4}|^2) = \mathrm{c}^2_{14} \sin^2 2\theta_{24} + \mathrm{s}^4_{24} \sin^2 2\theta_{14} \simeq \sin^2 2\theta_{24}\,,
\label{eq:MixAnglesSBL}
\end{align}
where it has been used that $U_{e4} = \mathrm{s}_{14}$ and $U_{\mu 4} = e^{-{\rm i} \delta_{24}} c_{14} s_{24}$. In the region dominated by the FD, i.e.~where $\Delta m_{31}^2 L/E = \mathcal{O}(1)$, 
an approximate appearance channel expression reads \cite{Boser:2019rta,Klop:2014ima}
\begin{align}
P^{\text{FD}}_{\mu e} &\simeq 4 \mathrm{s}^2_{13}\mathrm{s}^2_{23}\sin^2\Delta_{31} \nonumber\\
                &+ 8\mathrm{s}_{13}\mathrm{s}_{12}\mathrm{c}_{12}\mathrm{s}_{23}\mathrm{c}_{23}\sin\Delta_{21}\sin\Delta_{31}\cos(\Delta_{31}+\delta_{13}) \nonumber\\
                &+ 4\mathrm{s}_{13}\mathrm{s}_{14}\mathrm{s}_{24}\mathrm{s}_{23}\sin\Delta_{31}\sin(\Delta_{31}+\delta_{13}+\delta_{24})\,,
\label{eq:FmeFD}
\end{align}
where $\Delta_{ij} \equiv \Delta m^2_{ij} L/(4E)$.\footnote{In arriving at Eq.~(\ref{eq:FmeFD}), a series expansion in the small parameters $\Delta_{21}/\Delta_{31}$, $\mathrm{s}_{13}$, $\mathrm{s}_{14}$, and $\mathrm{s}_{24}$ is performed and it is assumed that $\Delta_{41} \to \infty$.} 
Note that in the last term of Eq.~(\ref{eq:FmeFD}), the relative sign between the phases are different compared to Refs.~\cite{Boser:2019rta,Klop:2014ima}, since we have used a different convention for the sterile CP-violating phases.

Similarly, using Eq.~(\ref{eq:probab}) in the so-called long-baseline limit (i.e.~$\Delta_{21} \to 0$ and $\Delta_{41} \to \infty$) or Eq.~(2.5) from Ref.~\cite{Kopp:2013vaa}, we derive a series expansion in $\mathrm{s}_{13}$, $\mathrm{s}_{14}$, and $\mathrm{s}_{24}$ for the disappearance channel formula in vacuum at the FD as
\begin{align}
P^{\text{FD}}_{\mu\mu} &\simeq 1 - \sin^2 2\theta_{23} \sin^2 \Delta_{31} \nonumber\\
		&+ 4 \mathrm{s}_{13}^2 \mathrm{s}_{23}^2 \cos 2\theta_{23} \sin^2 \Delta_{31} - \frac{1}{2} \mathrm{s}_{24}^2 \left( 3 + 2 \cos 4\theta_{23} \sin^2 \Delta_{31} + \cos 2\Delta_{31} \right) \nonumber\\
		&+ 4 \mathrm{s}_{13} \mathrm{s}_{14} \mathrm{s}_{24} \left( \sin 3\theta_{23} - \mathrm{s}_{23} \right) \cos(\delta_{13} + \delta_{24}) \sin^2 \Delta_{31}\,.
\label{eq:FmmFD}
\end{align}
Note that expressions~(\ref{eq:FmeFD}) and (\ref{eq:FmmFD}) are independent of $\Delta m^2_{41}$. The reason is that for the $L/E$ values which are relevant at an FD of long-baseline experiments (including super beams), oscillations are averaged out for the preferred values of $\Delta m_{41}^2$ by the LSND experiment~\cite{Aguilar:2001ty} or the MiniBooNE experiment~\cite{Aguilar-Arevalo:2018gpe} and global analyses~\cite{Dentler:2018sju}. Thus, long-baseline experiments are not sensitive to such values of $\Delta m^2_{41}$ if there is only an FD. Furthermore, we note that these expressions are independent of $\theta_{34}$ and $\delta_{34}$. However, as shown in Ref.~\cite{Gandhi:2015xza}, this simplification does not hold for the corresponding expression to Eq.~(\ref{eq:FmeFD}) in matter, since both $\theta_{34}$ and $\delta_{34}$ affect it significantly. Also, note that in the appearance channel probability, $\theta_{14}$, $\theta_{24}$, and $\delta_{24}$ appear in the third-order term in the small parameters $s_{13}$, $s_{14}$, and $s_{24}$ and $\delta_{13}$ appears in both the first-order and the third-order terms, whereas for the disappearance channel probability, $\theta_{14}$, $\delta_{13}$, and $\delta_{24}$ appear in the third-order term and $\theta_{24}$ appears in both the second-order and the third-order terms. Therefore, we understand that the constrains on $\theta_{24}$ coming from the disappearance channel will be stronger as compared to the constraint on $\theta_{14}$ from the same channel.

\section{Experimental setup and simulation details of ESSnuSB}
\label{sec3}

We perform our analysis using the GLoBES \cite{Huber:2004ka, Huber:2007ji} software to simulate ESSnuSB. 
We explicitly simulate an ND and an FD in order to reduce systematic uncertainties~\cite{Baussan:2013zcy}. The FD is a 1~Mt MEMPHYS-like water-\cnv\ detector~\cite{Agostino:2012fd} located at a distance of 540~km from the source, while the ND is assumed to have the same efficiency and background rejection capabilities as the FD~\cite{Agostino:2012fd} with a fiducial volume of 0.1~kt and placed at a distance of 0.5~km from the source. We consider a beam power of 5~MW with 2.5~GeV protons capable of producing $2.7\times 10^{23}$ protons on target per year. The fluxes are simulated explicitly at 1~km for the ND~\cite{Blennow:2014fqa} as well as at 100~km (and consequently rescaled) for the longer baseline of the FD~\cite{Baussan:2013zcy}. We assume a total 10~years of running time divided into three options, which are 2+8, 5+5, and 8+2, where ``a+b'' implies, ``a'' years running in neutrino mode and ``b'' years running in antineutrino mode. Throughout the simulations, we use the same treatment of systematic errors as in Ref.~\cite{Coloma:2012ji}. In particular, we use the systematic uncertainties from Tab.~\ref{Tab:Systematics}, which correspond to the dubbed ``Default'' systematics from Ref.~\cite{Coloma:2012ji}.
\begin{table}
\centering
\begin{tabular} {| c | c |}
\hline
Systematics &  Default \\
\hline
Fiducial volume ND & 0.5~\% \\
Fiducial volume FD & 2.5~\% \\
Flux error $\nu$ & 7.5~\% \\
Flux error $\bar{\nu}$ & 15~\% \\
Neutral current background & 7.5~\% \\
Cross section $\times$ eff. QE & 15~\% \\
Ratio $\nu_e/\nu_{\mu}$ QE & 11~\% \\
\hline
\end{tabular}
\caption{Systematic uncertainties for a super beam as described in Ref.~\cite{Coloma:2012ji} for the ``Default'' scenario.}
\label{Tab:Systematics}
\end{table}
For the FD only analysis, we consider an overall 8~\% systematic error in the signal and a 10~\% one error in the background. We consider both the appearance and the disappearance channels in our analysis.

We estimate the statistical $\chi^2$ function using
\begin{equation}
\chi^2_{{\rm stat}} = 2 \sum_{i=1}^n \bigg[ N^{{\rm test}}_i - N^{{\rm true}}_i - N^{{\rm true}}_i \log\bigg(\frac{N^{{\rm test}}_i}{N^{{\rm true}}_i}\bigg) \bigg]\,,
\end{equation}
where $n$ is the number of energy bins, $N^{{\rm true}}$ is the number of true events, and $N^{{\rm test}}$ is the number of test events and incorporate the systematics by the method of pulls. We also include Gaussian priors on all standard neutrino oscillation parameters with their $1\sigma$ errors to the $\chi^2$ function, except on $\theta_{23}$, $\Delta m_{31}^2$, and $\delta_{13}$, for which we include Gaussian priors on $\sin^2{2\theta_{23}}$ and $|\Delta m_{31}^2|$, and no prior at all on $\delta_{13}$. We do not consider any priors on the sterile mixing parameters as well when marginalizing over them.

\section{Simulation results}
\label{sec4}

In this section, we will present our results. First, we will present results at the probability level, which will help to understand the nature of neutrino oscillations in the 3+1 scenario at the FD and the ND. Then, we will present the sensitivity of ESSnuSB to constrain the sterile mixing parameters and discuss how the sensitivity of ESSnuSB compares to the ones of other long-baseline neutrino oscillation experiments. Finally, we will discuss the CP violation sensitivity of ESSnuSB in the presence of light sterile neutrinos. For the presentation of all results, we assume the best-fit values of the parameters from Ref.~\cite{Esteban:2018azc} assuming NO, which are currently preferred and summarized in Tab.~\ref{Tab:Param}.
\begin{table}
	\centering
	\setlength{\extrarowheight}{0.1cm}
	\begin{tabular}{|c|c|c|}
		\hline
		 Parameter & Best-fit value & 1$\sigma$ error\\
		\hline
		$\theta_{12}$ & $33.82^\circ$ & 2.3~\%\\
		$\theta_{13}$ & $8.60^\circ$ & 1.5~\%\\
		$\theta_{23}$ & $48.6^\circ$ & 2.8~\%\\
		$\delta_{13}$ & $-139^\circ$ & N/A\\
		$\Delta m^2_{21}$ & $7.39\times10^{-5}\mathrm{eV}^2$ & 2.8~\% \\
		$\Delta m^2_{31}$ & $2.53\times10^{-3}\mathrm{eV}^2$ & 1.2~\%\\
		\hline
	\end{tabular}
	\caption{Best-fit values of the neutrino oscillation parameters for the standard three-flavor scenario adopted from Ref.~\cite{Esteban:2018azc}.}
	\label{Tab:Param}
\end{table}

\subsection{Discussion at the probability level}
\label{Sec:ProbLevel}

\begin{figure}
\includegraphics[width=0.6\textwidth]{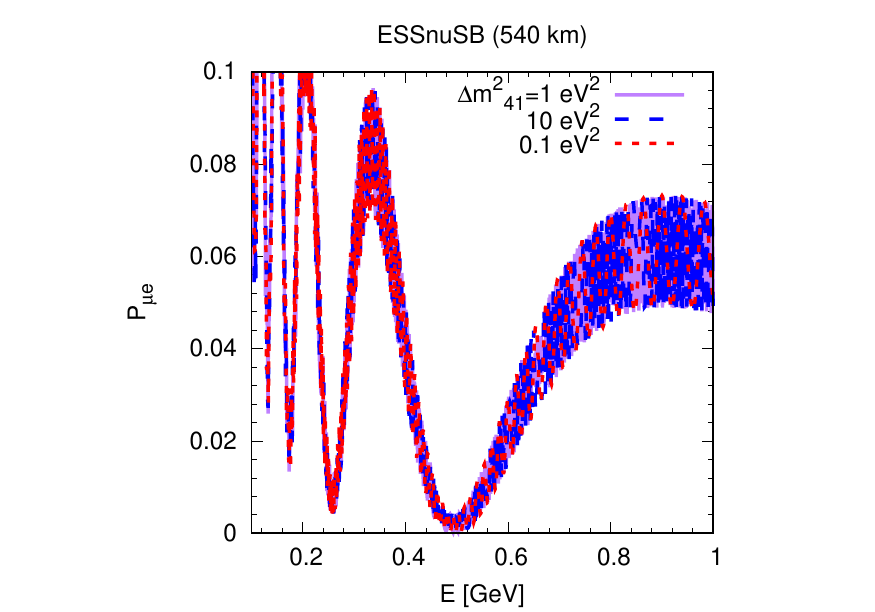} 
\hspace{-0.7 in}
\includegraphics[width=0.6\textwidth]{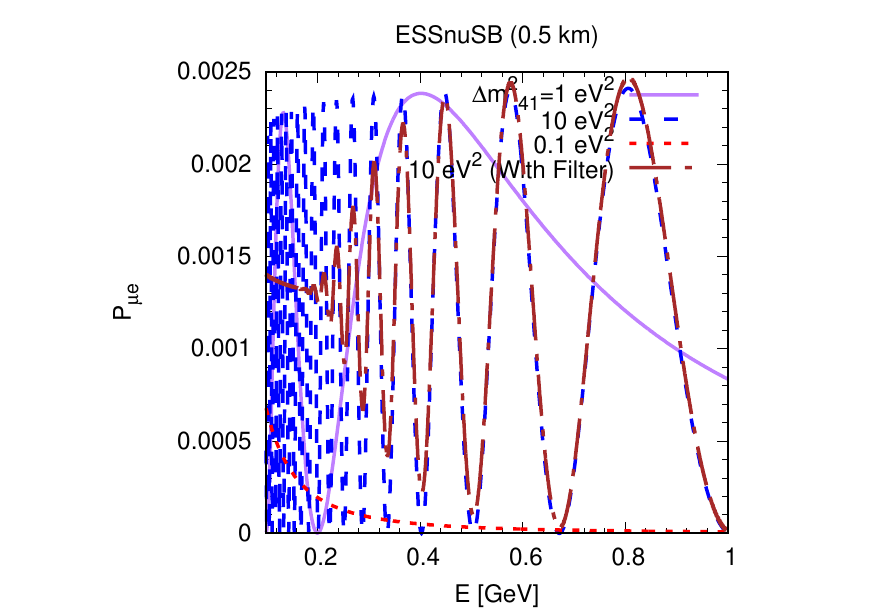} 
\caption{Appearance channel oscillation probability for neutrinos as a function of the neutrino energy $E$. The left panel is for ESSnuSB with a baseline length of 540~km, whereas the right panel is for ESSnuSB with a baseline length of 0.5~km. The solid purple curve corresponds to $\Delta m_{41}^2=1$~eV$^2$, the long-dashed blue one to $\Delta m_{41}^2=10$~eV$^2$, and finally, the dashed green one to $\Delta m_{41}^2=0.1$~eV$^2$. In the right-panel, the dot-dashed maroon curve corresponds to the oscillation probability convoluted with the Gaussian filter to take care of the fast oscillations.}
\label{fig:Probability}
\end{figure}

In Fig.~\ref{fig:Probability}, we show the appearance channel oscillation probability for neutrinos as a function of $E$ in the presence of sterile neutrinos with different values of $\Delta m_{41}^2$.
The left panel is for the baseline length of 540~km, which corresponds to the distance to the FD, whereas the right panel is for the baseline length of 0.5~km, corresponding to the distance to the ND. These panels are generated for the best-fit values of the neutrino oscillation parameters for the standard three-flavor scenario given in Tab.~\ref{Tab:Param}, taken from Ref.~\cite{Esteban:2018azc}.
For the sterile mixing parameters, we choose $\sin^2\theta_{14} = \sin^2\theta_{24} =  0.025$ \cite{Diaz:2019fwt}. Furthermore, we set the values of the sterile CP-violating phases to zero, i.e.~$\delta_{24} = \delta_{34} = 0^\circ$. Each panel has three curves corresponding to three different values of $\Delta m^2_{41}$, which are 1~eV$^2$, 10~eV$^2$, and 0.01~eV$^2$, respectively. From the left panel, we see that the probability for all three values of $\Delta m^2_{41}$ are equivalent and there is no distinguishable behavior in the oscillation pattern, since at this value of $L/E$, the oscillations are averaged out as can be seen in Eq.~(\ref{eq:FmeFD}). As expected, if we perform an analysis with an FD only, it will not be possible to probe such large values of $\Delta m^2_{41}$. On the other hand, from the right panel, we observe that for an ND, the oscillation probability is sensitive to different values of $\Delta m^2_{41}$. For $\Delta m^2_{41} = 0.1 $~eV$^2$, we note that the oscillations have not fully developed, while for $\Delta m^2_{41} = 10 $~eV$^2$, the oscillations tend to become rapid compared to the 0.1~GeV energy resolution of a water-\cnv\ detector. In order to perform the sensitivity studies in GLoBES, we use the built-in filter that convolutes the oscillation probability with a Gaussian with width approximately that of the energy bin, such that the sensitivity obtained is not spurious due to the fast oscillations.  The use of the filter results in the oscillation probability shown by the dot-dashed maroon curve in the right panel. Finally, for $\Delta m^2_{41} = 1$~eV$^2$, we note that the first oscillation maximum is being fully probed at the detector. Therefore, we understand that an ND is ideal to measure the oscillations governed by a light sterile neutrino of mass around 1~eV$^2$, and thus, we expect to obtain good sensitivity to the sterile mixing parameters if we perform a combined analysis with both FD and ND.

\subsection{Bounds on sterile mixing parameters}
\label{Bounds_t14vst24}

\begin{figure}
\begin{center}
\includegraphics[width=0.65\textwidth]{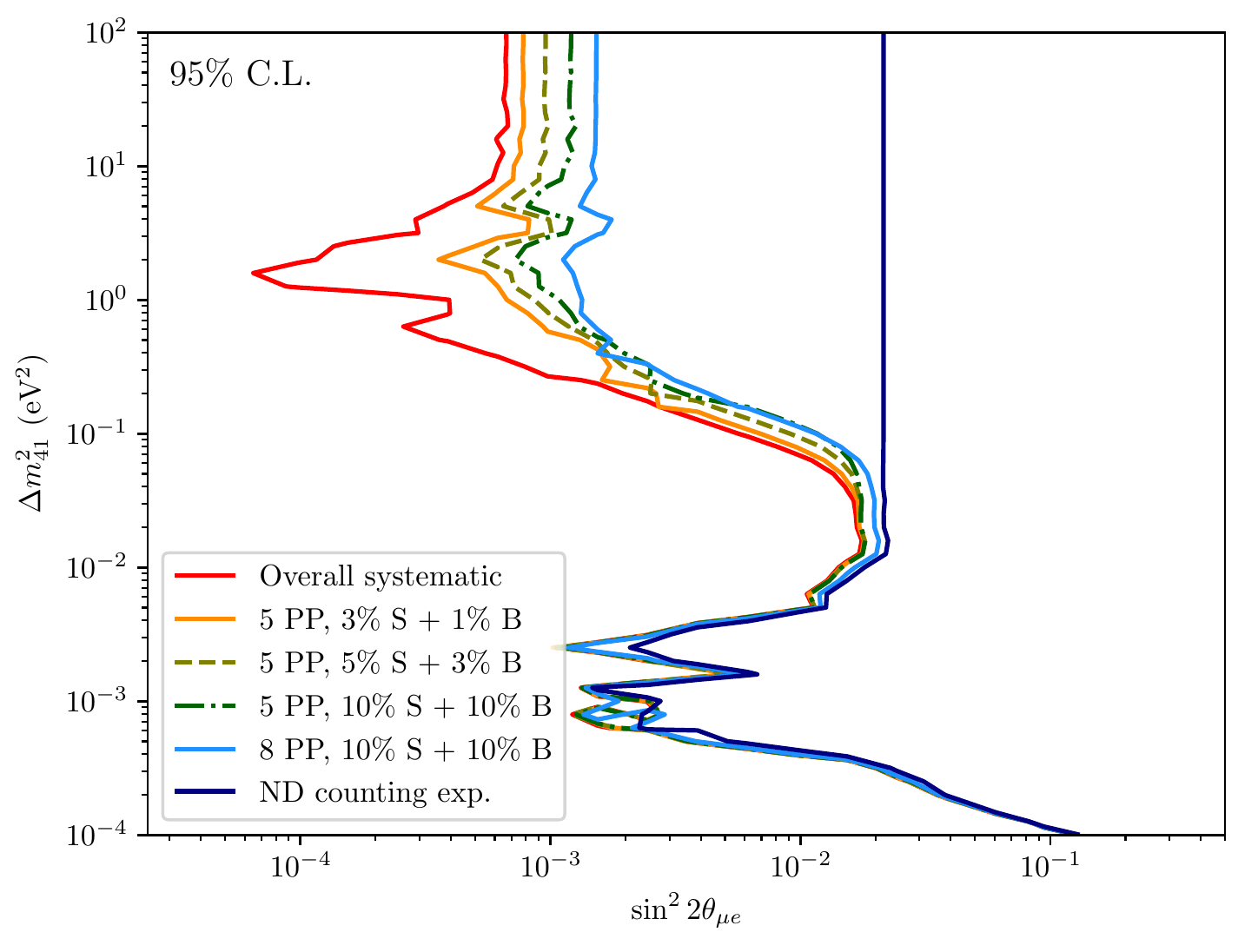} 
\caption{Plot in the $\sin^2{\theta_{\mu e}}$ -- $\Delta m^2_{41}$ plane: Sensitivity to the sterile mixing parameters of interest necessary to describe the results of the LSND experiment for different choices of systematic errors. The red curve corresponds to having only overall systematics, the dark blue one accounts to having no spectral information in the near detector, while the rest of the curves correspond to introducing different levels of shape systematic uncertainties in the signal (S) and the background (B) channels for different number of pivot points (PP) where the systematics is allowed to freely vary.}
\label{fig:ESSnuSB_sens_LSND}
\end{center}
\end{figure}

Next, we discuss how well ESSnuSB can put bounds on the sterile mixing parameters. As noted in the right panel of Fig.~\ref{fig:Probability}, for given values of $\Delta m_{41}^2$ we have very fast oscillations compared to the energy resolution of the detector, which would be averaged out. In Fig.~\ref{fig:ESSnuSB_sens_LSND}, we show the sensitivity at 95~\%~C.L. to sterile neutrinos that could describe the electron neutrino appearance measured in the former LSND experiment~\cite{Aguilar:2001ty} for different values of $\Delta m_{41}^2$ and $\sin^2{2\theta_{\mu e}}$ as defined in Eq.~(\ref{eq:MixAnglesSBL}). We marginalize over $\theta_{14}$ and $\theta_{24}$ and keep the rest of sterile parameters fixed to zero in the test values. The splitting of the running time is $5+5$ years in neutrino and antineutrino mode, respectively.

As we can see from Fig.~\ref{fig:ESSnuSB_sens_LSND}, there are two well-defined sensitivity regions for the different values of $\Delta m_{41}^2$. The FD-dominated region for $\Delta m_{41}^2\sim 10^{-3}$~eV$^2$ (i.e.~near values of $\Delta m_{31}^2$), where the sterile oscillations are fully developed at the FD, and the ND-dominated region for $\Delta m_{41}^2\sim 1$~eV$^2$, where the first sterile oscillation maximum is developed at the ND, as shown in Fig.~\ref{fig:Probability}. Although the detector has only an energy resolution of 0.1~GeV, this is enough to reconstruct the shape of the event rate such that it cannot be mimicked by overall normalization systematics (solid red curve in Fig.~\ref{fig:ESSnuSB_sens_LSND}). Thus, the sensitivity around 1~eV$^2$ is at the order of $\sin^2{2\theta_{\mu e}}\sim10^{-4}$. On the other hand, if the ND had no energy resolution, i.e.~being a ``counting experiment'' (solid dark blue curve in Fig.~\ref{fig:ESSnuSB_sens_LSND}), then there would be no shape information to reconstruct the oscillations from sterile neutrinos, and therefore, overall systematics could hide the presence of sterile neutrinos, dramatically reducing the sensitivity down to $\sin^2{2\theta_{\mu e}}\sim 10^{-2}$. 

It is interesting to study the impact of systematic uncertainties which affect the shape of the event rate. This kind of systematics could arise from uncertainties in the neutrino flux or the cross section, which depend on the particular energy of the neutrinos affecting different experimental energy bins independently. Such systematic uncertainties are implemented in the simulations, using different sizes of the systematics for the signal (S) and the background (B) channels and also different energies at which the systematics is allowed to freely vary (5 pivot points for the solid orange, dashed olive, and dark dot-dashed green curves or 8 pivot points for the solid light blue curve in Fig.~\ref{fig:ESSnuSB_sens_LSND}). As can be seen, already when introducing small systematics in the shape of the event rate, the sensitivity around $\Delta m_{41}^2$ is reduced around an order of magnitude or even more depending on the size of this systematics. Although a detailed study beyond the scope of this work would be necessary to appropriately adress the adequate size and impact of these uncertainties, it is clear that the energy resolution of the ND is crucial to have as large sensitivity as possible to sterile neutrinos.

\begin{figure}
\includegraphics[width=0.42\textwidth]{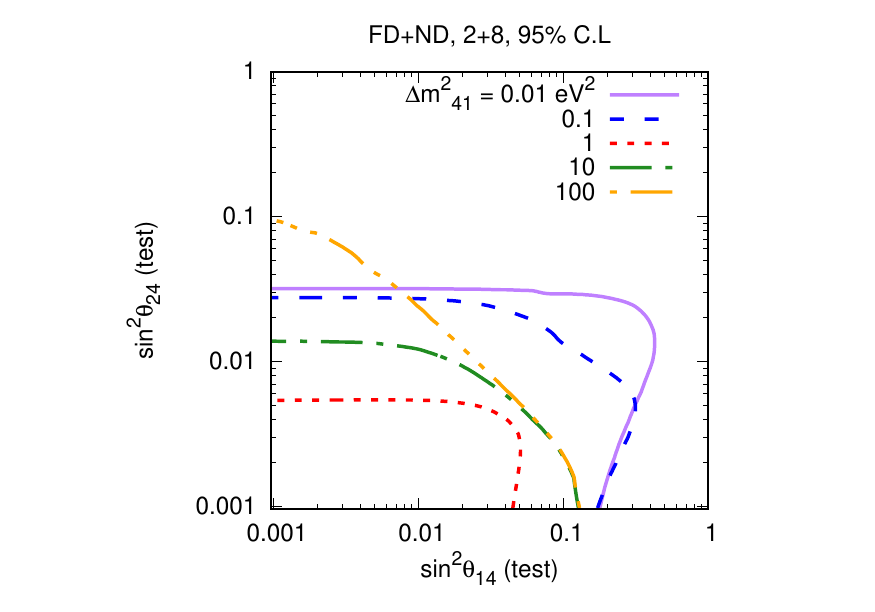} 
\hspace{-0.9 in}
\includegraphics[width=0.42\textwidth]{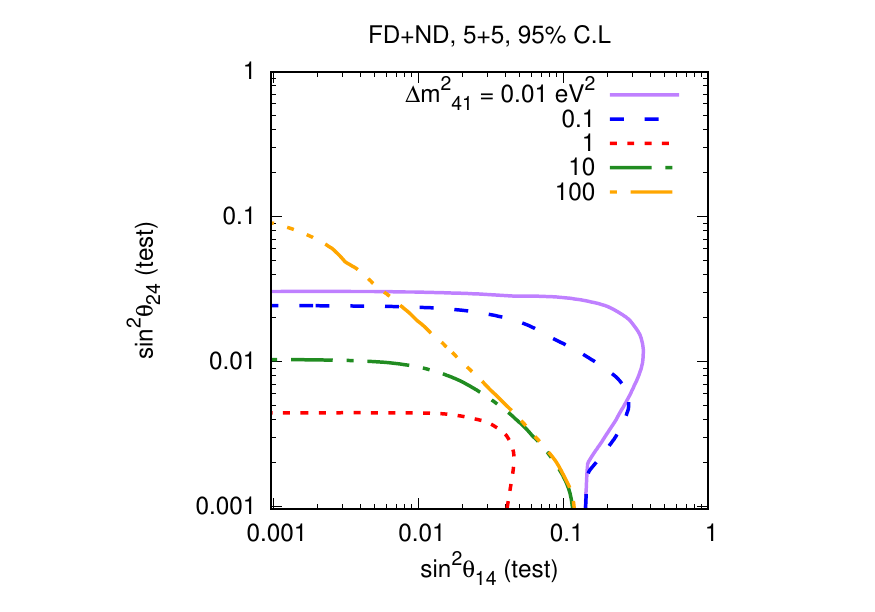} 
\hspace{-0.9 in}
\includegraphics[width=0.42\textwidth]{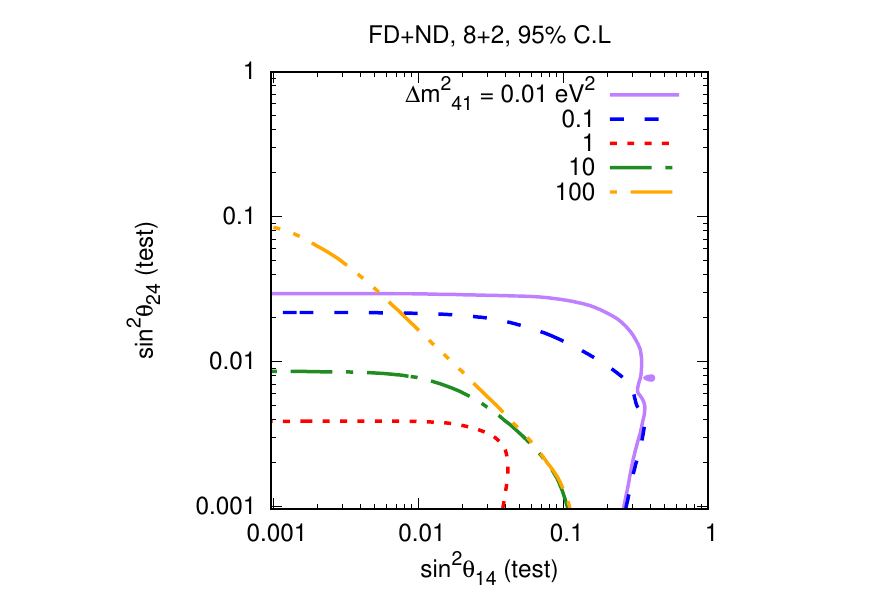} \\
\includegraphics[width=0.42\textwidth]{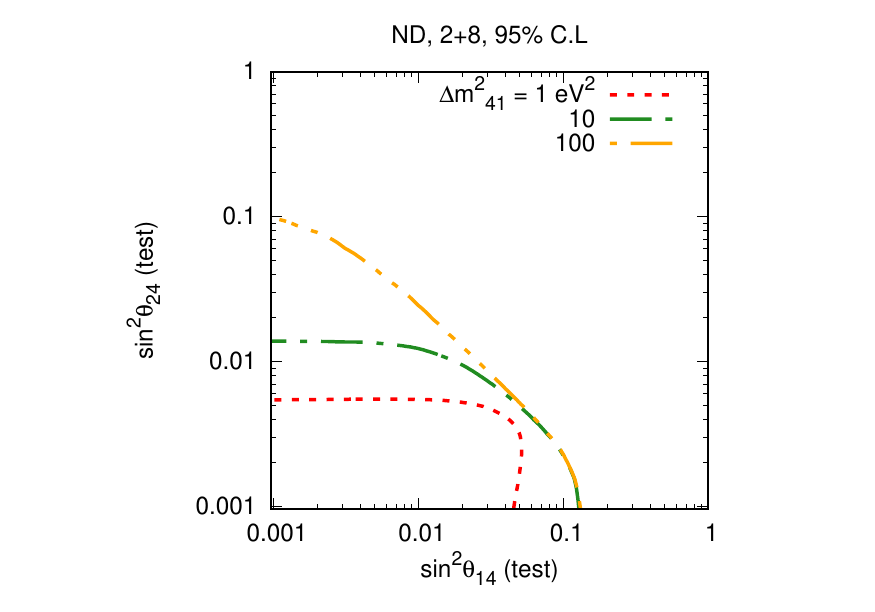} 
\hspace{-0.9 in}
\includegraphics[width=0.42\textwidth]{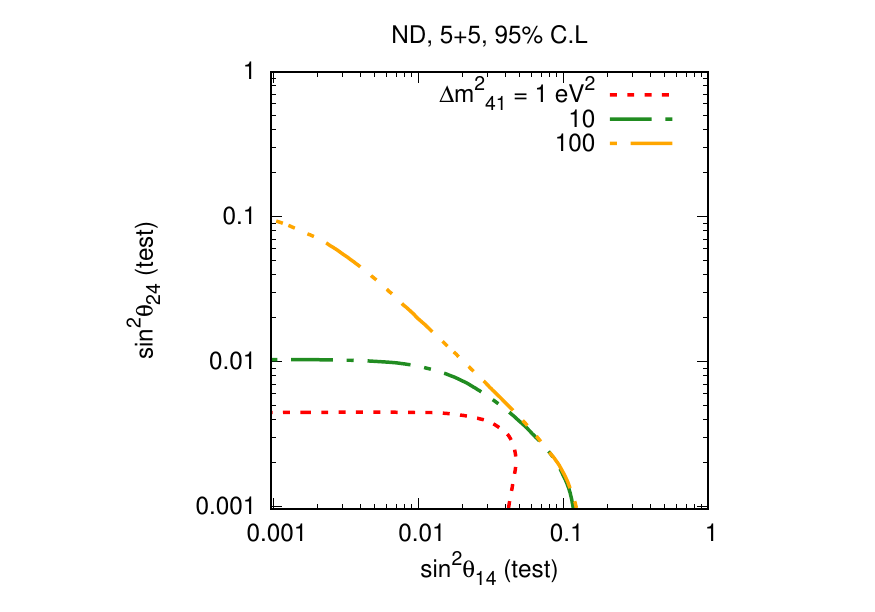} 
\hspace{-0.9 in}
\includegraphics[width=0.42\textwidth]{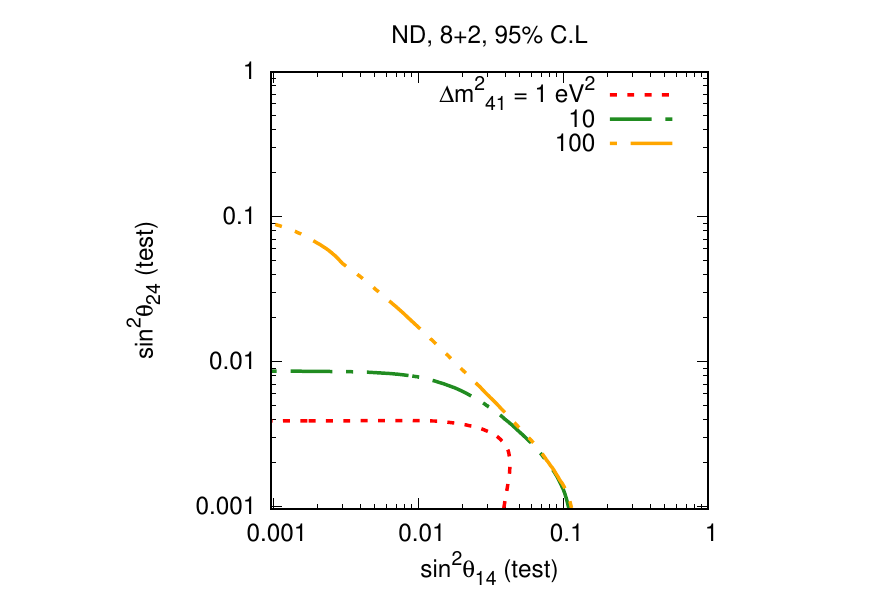}  \\
\includegraphics[width=0.42\textwidth]{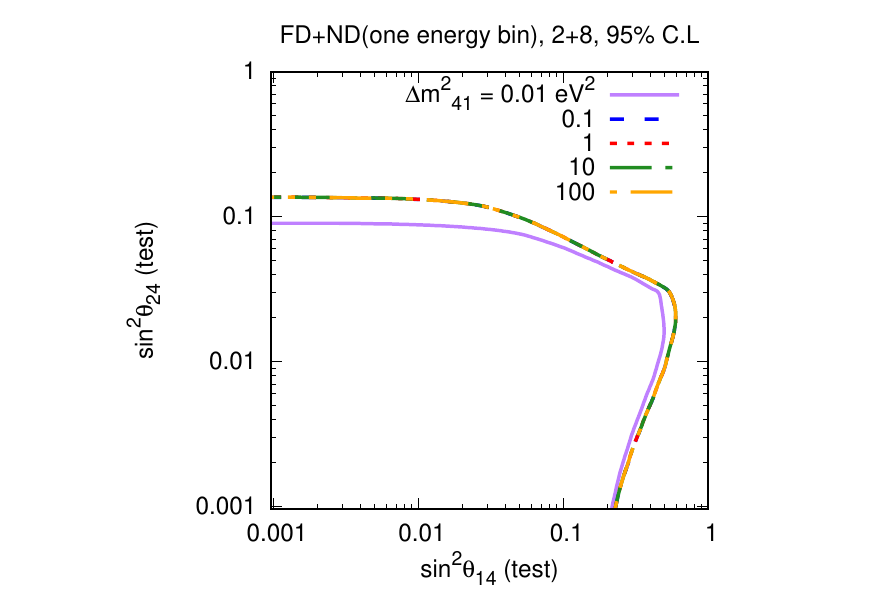} 
\hspace{-0.9 in}
\includegraphics[width=0.42\textwidth]{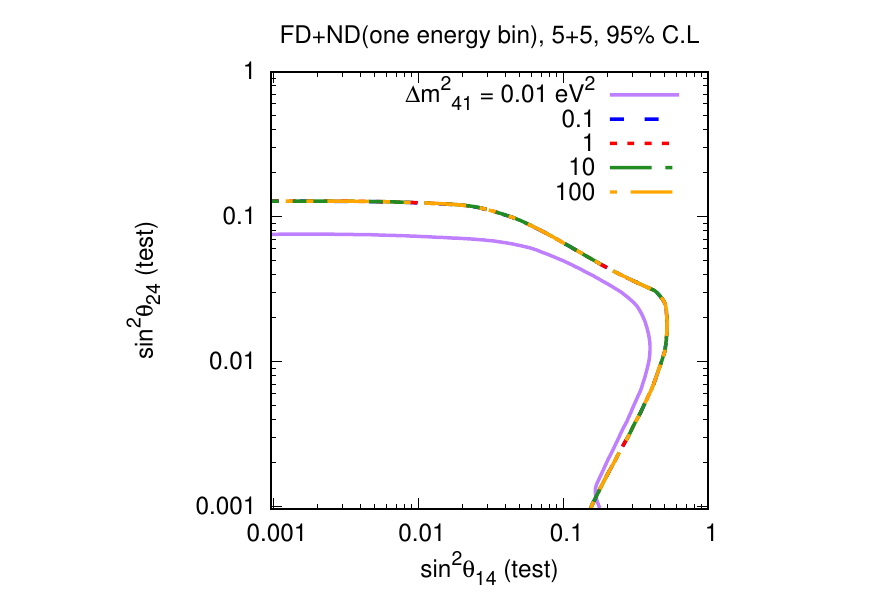} 
\hspace{-0.9 in}
\includegraphics[width=0.42\textwidth]{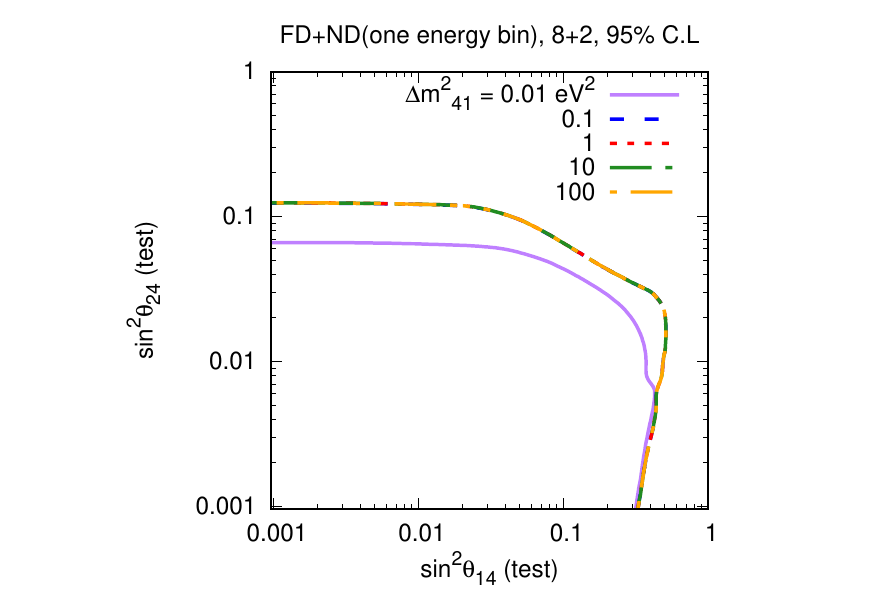}  \\
\caption{Bounds (95~\% C.L.) on sterile mixing parameters in the $\sin^2\theta_{14}$(test) -- $\sin^2\theta_{24}$(test) plane for ESSnuSB.}
\label{fig:Sens_t14vst24}
\end{figure}

In Fig.~\ref{fig:Sens_t14vst24}, we present the sensitivity at 95~\%~C.L. of ESSnuSB to the sterile mixing angles $\theta_{14}$ and $\theta_{24}$ for particular values of $\Delta m_{41}^2$. We marginalize over all standard neutrino oscillation parameters as well as the sterile parameters not explicitly shown in the plots of Fig.~\ref{fig:Sens_t14vst24}, namely $\theta_{34}$, $\delta_{14}$, and $\delta_{34}$. 
The left, middle, and right columns are for ESSnuSB running options of 2+8, 5+5, and 8+2, respectively. In the top row, we present the results for combined FD+ND with the systematics from Tab.~\ref{Tab:Systematics}, and in the middle row, we give the results for an ND only. Finally, in the bottom row, in order to estimate the sensitivity for the FD without any ND, we present the results for a detector configuration of FD+ND without any energy information on the ND such that sterile oscillations cannot be observed at the ND. In other words, we consider only one energy bin for the ND, acting as a counting experiment, which corresponds to the solid dark blue curve in Fig.~\ref{fig:ESSnuSB_sens_LSND}. We adopt this method so that the systematic uncertainties in every row in Fig.~\ref{fig:Sens_t14vst24} are exactly the same ones among the different cases and we can fairly compare every panel of this figure.

From the top row, we see that for $\Delta m^2_{41} = 0.01$~eV$^2$, the bounds are weak as for this value of $\Delta m^2_{41}$, the oscillations have not yet developed for the ND, and therefore, the existing sensitivity comes from the FD, where the oscillations are averaged out. 
Increasing the value of $\Delta m^2_{41}$, the oscillations becomes more developed in the ND and we obtain the best bound for $\Delta m^2_{41} = 1$~eV$^2$, since for this value of $\Delta m^2_{41}$, the oscillations are fully developed at the ND, which can be observed in Fig.~\ref{fig:Probability}. Increasing the value of $\Delta m^2_{41}$ further, the oscillations tend to become averaged out and again the sensitivity decreases. In the middle row, as the oscillations for $\Delta m^2_{41} = 0.01$~eV$^2$ and 0.1~eV$^2$ have not developed at the ND, we only present results for 1~eV$^2$, 10~eV$^2$, and 100~eV$^2$. From these curves, we note that the sensitivity is very similar to that of FD+ND. From this, we understand that for these values of  $\Delta m^2_{41}$ even for FD+ND, the main sensitivity comes from the ND and including data from the FD does not help much as it becomes systematically dominant with further addition of statistics. From the bottom row, we note that when there is no spectral information on the ND, the sensitivity for  $\Delta m^2_{41} = 1$~eV$^2$ is lost and for all values of $\Delta m^2_{41}$ we obtain similar sensitivities. The reason is that now the sensitivity comes from the FD, where the oscillations due to the sterile neutrinos are averaged out and the sensitivity due to different values of  $\Delta m^2_{41}$ is lost. From the plots, we note that the bounds on $\theta_{24}$ are slightly better than the bounds on $\theta_{14}$. The reason is that the extra $\theta_{24}$ sensitivity stems from the second-order term in the disappearance channel probability [see Eq.~(\ref{eq:FmmFD})]. Regarding the three different running options, we note that with the increment of the fraction of the neutrino running, the sensitivity becomes slightly better.

\subsection{Comparison with other long-baseline experiments}
\label{sec:comparison}

\begin{figure}
\begin{center}
\includegraphics[width=0.7\textwidth]{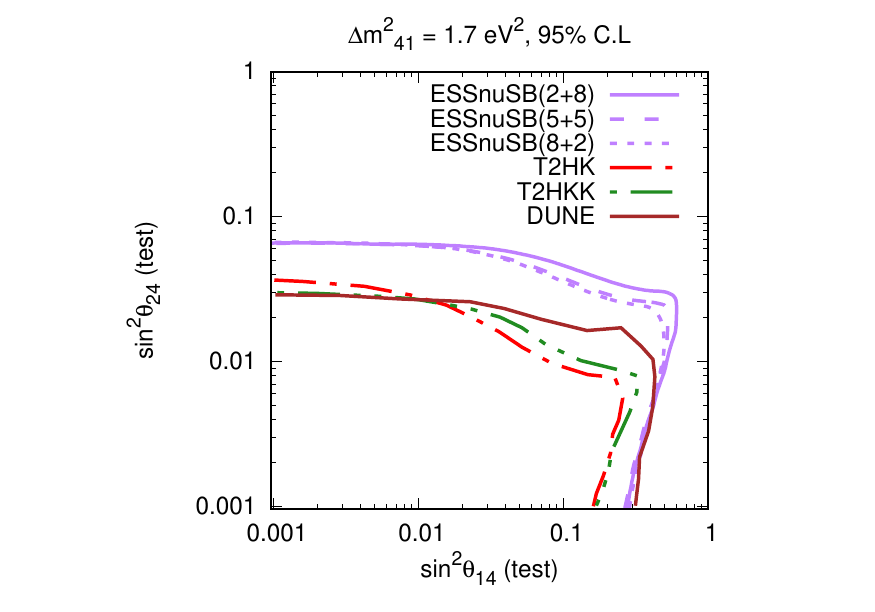} 
\caption{Comparison of ESSnuSB with other long-baseline neutrino oscillation experiments. This plot is for an FD only with 8~\% overall systematic error in the signal and 10~\% overall systematic error in the background.}
\label{fig3}
\end{center}
\end{figure}

In this subsection, we will compare the sensitivity of ESSnuSB to the sterile mixing parameters with other future long-baseline neutrino oscillation experiments, namely T2HK \cite{Abe:2016ero}, T2HKK \cite{Abe:2016ero}, and DUNE \cite{Acciarri:2015uup}. T2HK will be the upgrade of the T2K experiment, where the detector will be upgraded to two water-\cnv\ tanks of 187~kt each near the existing Kamioka site. The proposal of the T2HKK project is to move one of the water-\cnv\ tanks to Korea at a baseline length of 1100~km. DUNE is the long-baseline neutrino oscillation program of Fermilab having a baseline length of 1300~km. In Fig.~\ref{fig3}, we present the sensitivity of ESSnuSB in the $\sin^2\theta_{14}$(test) -- $\sin^2\theta_{24}$(test) plane at 95~\% C.L.~for three combinations of the neutrino-antineutrino running ratio (the same as that of Fig.~\ref{fig:Sens_t14vst24}) along with the sensitivities of T2HK, T2HKK, and DUNE as obtained in Ref.~\cite{Choubey:2017ppj}.
As the sensitivities of T2HK, T2HKK, and DUNE are obtained assuming an FD only, we present our results of ESSnuSB considering an FD only and assuming an overall systematics of 8~\% in the signal and 10~\% in the background. We have checked that this configuration of ESSnuSB exhibits similar sensitivity as the FD+ND configuration with correlated systematics for the search of CP violation, for which the experiment was originally designed~\cite{Baussan:2013zcy}. For these results, we choose the same true parameter values as used in Ref.~\cite{Choubey:2017ppj}, which are $\theta_{12} = 33.56^\circ$, $\theta_{13} = 8.46^\circ$, $\theta_{23} = 45^\circ$, $\Delta m^2_{21} = 7.5 \times 10^{-5}$ eV$^2$, $\Delta m^2_{31} = 2.5 \times 10^{-3}$ eV$^2$, and $\delta_{13} = -90^\circ$. Following Ref.~\cite{Choubey:2017ppj}, we marginalize only over the parameters $\delta_{13}$, $\theta_{14}$, $\theta_{24}$, $\theta_{34}$, $\delta_{24}$, and $\delta_{34}$. We use the GLoBES minimizer and assume no priors. The value of $\Delta m^2_{41}$ is 1.7~eV$^2$. From the plot, we see that the sensitivity of ESSnuSB is slightly worse than the other experiments in most of the region of the parameter space, given that an FD only cannot probe the sterile oscillation for such value of $\Delta m_{41}^2$. Note however, that from the first row of panels in Fig.~\ref{fig:Sens_t14vst24} it is expected that the ESSnuSB sensitivity is considerably improved in the presence of an ND. We also note that around $\sin^2 \theta_{14}$(test) $= 0.3$, the sensitivity of ESSnuSB is comparable with that of DUNE. As mentioned earlier, the ESSnuSB run time of 8+2 year gives slightly better sensitivity than the other two configurations. 

\subsection{CP sensitivity in presence of light sterile neutrinos}
\label{sec:CPsensitivity}

\begin{figure}
\includegraphics[width=0.42\textwidth]{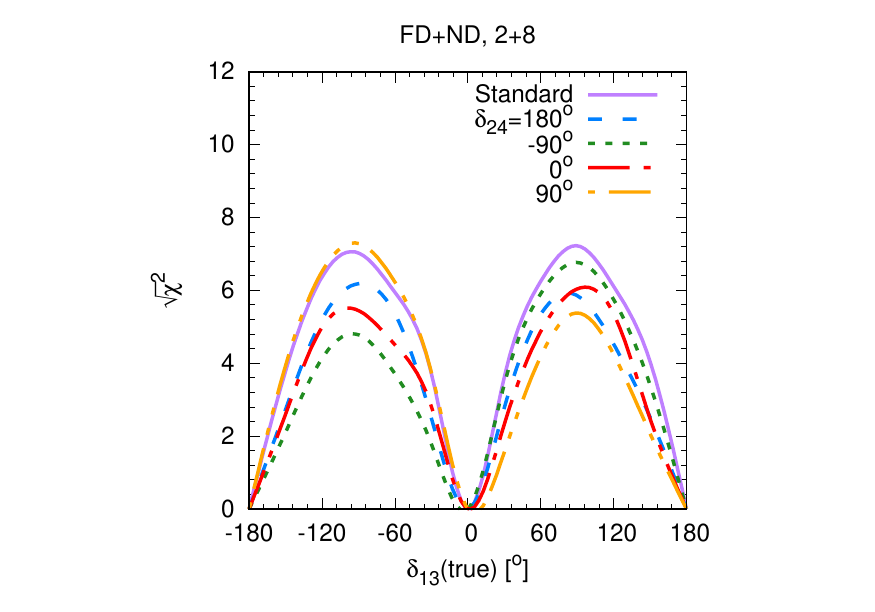} 
\hspace{-0.9 in}
\includegraphics[width=0.42\textwidth]{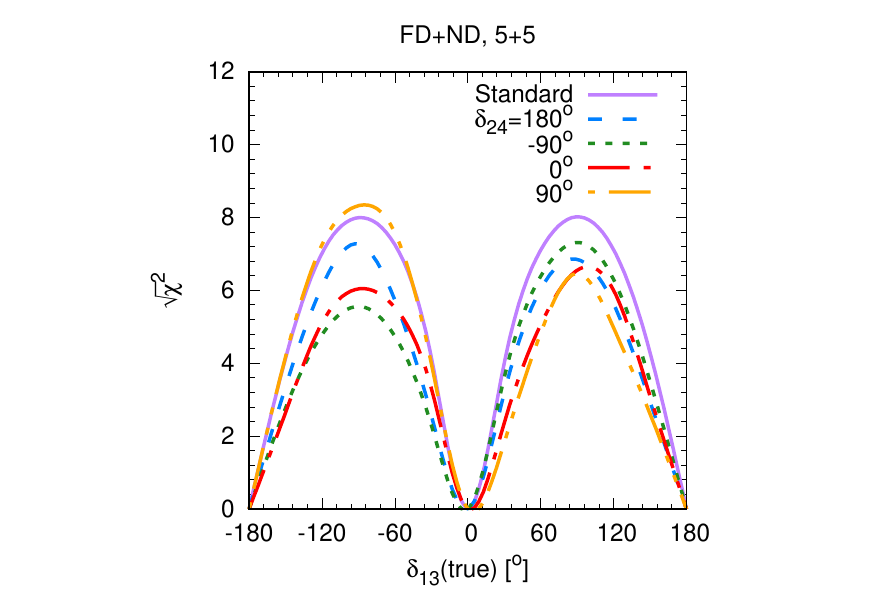} 
\hspace{-0.9 in}
\includegraphics[width=0.42\textwidth]{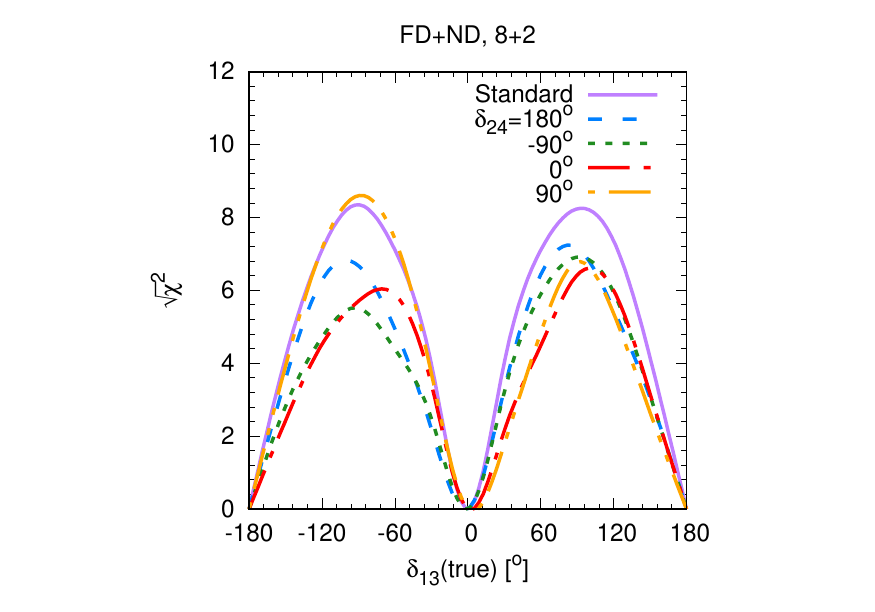} \\
\includegraphics[width=0.42\textwidth]{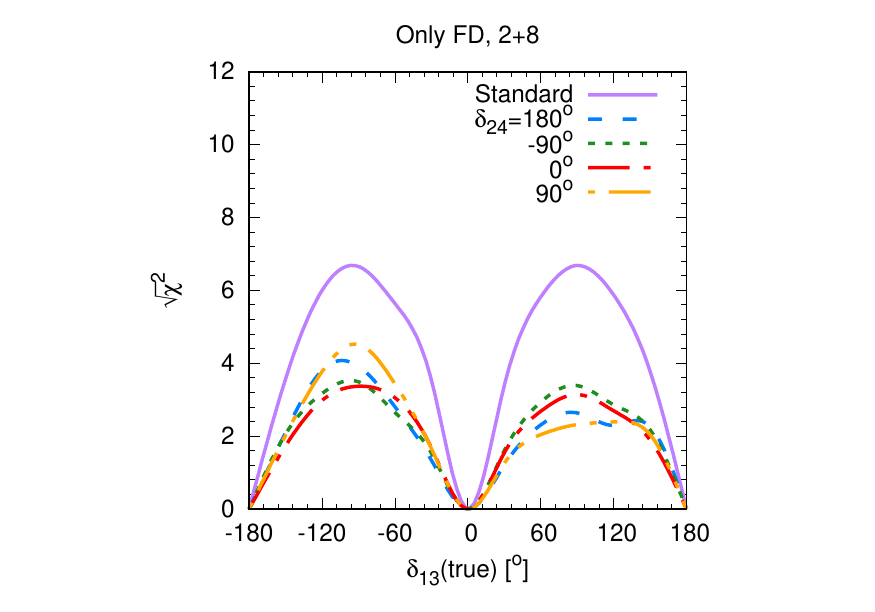} 
\hspace{-0.9 in}
\includegraphics[width=0.42\textwidth]{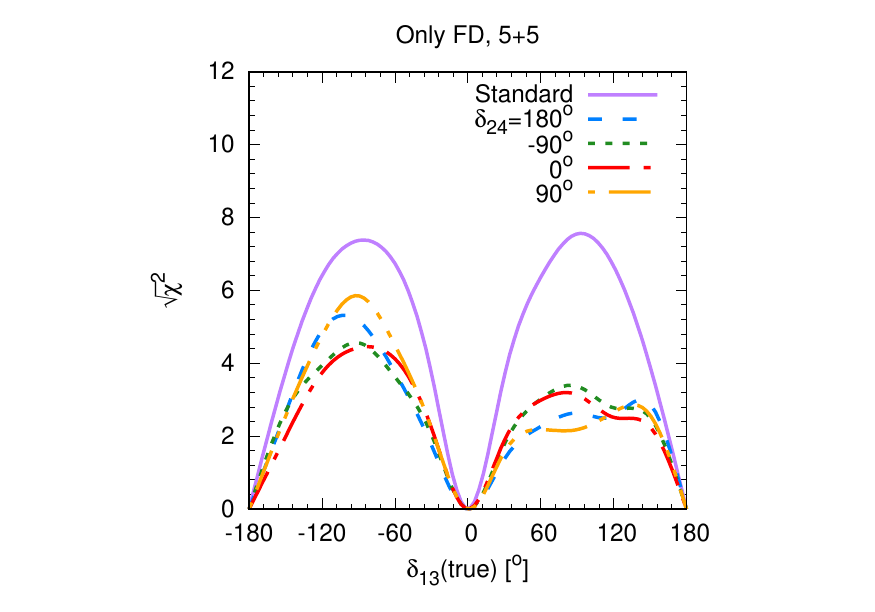} 
\hspace{-0.9 in}
\includegraphics[width=0.42\textwidth]{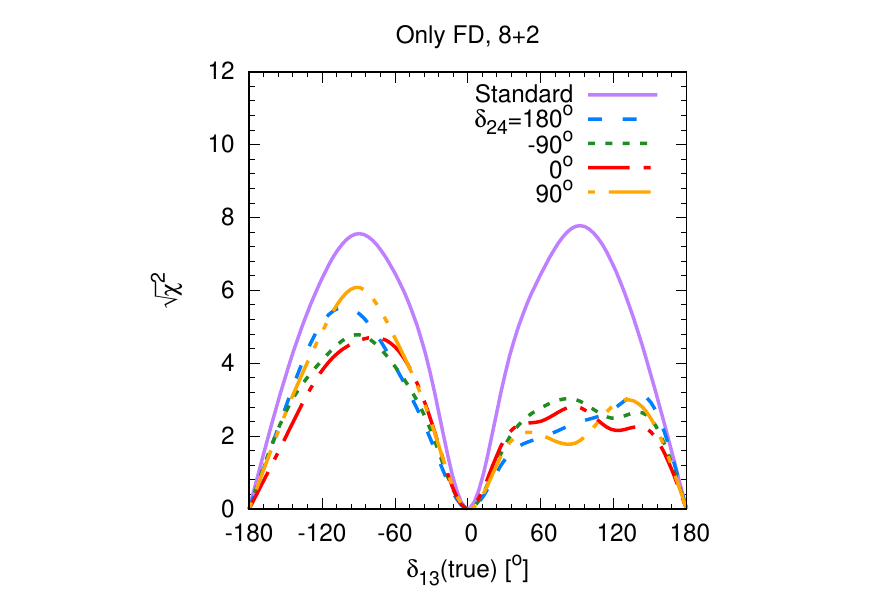}  \\
\caption{CP violation sensitivity of ESSnuSB for four different values of true $\delta_{24}$ as a function of true $\delta_{13}$.}
\label{fig:CP_violation}
\end{figure}

In this subsection, we will discuss the CP violation sensitivity of ESSnuSB assuming the existence of a light sterile neutrino with $\Delta m^2_{41} = 1$ eV$^2$. CP violation sensitivity of an experiment is defined by its capability to distinguish a value of the CP-violating phase other than $0^\circ$ and $180^\circ$. In Fig.~\ref{fig:CP_violation}, we present the CP violation sensitivity of ESSnuSB due to $\delta_{13}$ for four different values of true $\delta_{24}$.
The true values of the standard neutrino oscillation parameters are the ones given in Tab.~\ref{Tab:Param}. For the sterile mixing parameters, we assume $\sin^2\theta_{14} = \sin^2\theta_{24} = 0.025$ \cite{Diaz:2019fwt}, $\theta_{34} = \delta_{34} = 0^\circ$, and $\Delta m^2_{41} = 1$~eV$^2$. We marginalize over all standard parameters with $\delta_{13}$ minimized over $0^\circ$ and $180^\circ$ only and also over the sterile parameters $\theta_{14}$, $\theta_{24}$, $\theta_{34}$, $\delta_{24}$, and $\delta_{34}$. In these plots, we minimize over the neutrino mass ordering, i.e.~we choose the minimum of the two calculated values of the $\chi^2$ function for both test NO and test IO, and both octants.
In the top row, we present the sensitivity for FD+ND, and in the bottom row, we present the sensitivities for an FD only with 8~\% (10~\%) overall systematics in signal (background). The left, middle, and right panels are for the ESSnuSB configurations of 2+8, 5+5, and 8+2 neutrino-antineutrino running times, respectively. The figures in the bottom panels are consistent with the results from Ref.~\cite{Agarwalla:2019zex}. From these plots, we see that in the absence of an ND, the CP violation sensitivity in the presence of sterile neutrinos is lower as compared to the case when we include an ND along with an FD. However, when we include the ND, the sterile mixing parameters are constrained due to oscillations of sterile neutrinos at the ND, as discussed in Section~\ref{Bounds_t14vst24}, which is not the case for an FD only, where oscillations of sterile neutrinos are averaged out and the constraints on the sterile mixing parameters are weaker. Therefore, the conclusion obtained in Ref.~\cite{Agarwalla:2019zex} that the CP violation sensitivity deteriorates substantially in the presence of sterile neutrinos can be drastically altered if there is an ND. We have checked that for the case of an FD only, the sensitivity is affected due to the marginalization of $\theta_{14}$ (if $\theta_{14}$ is fixed, then the value of the $\chi^2$ function becomes very high). We also note that the sensitivity of $\delta_{13}$ around $90^\circ$ for an FD only is affected more when we increase the neutrino running time. For $\delta_{13}\sim 90^{\circ}$, the oscillation probability for antineutrinos is enhanced with respect to neutrinos, but the running time for antineutrinos is reduced, so statistics become reduced as the neutrino running time is increased. For the combined FD and ND, we note an asymmetry in the curve for $\delta_{24} = -90^\circ$ and $\delta_{24} = 90^\circ$. For $\delta_{24} = -90^\circ$ ($90^\circ$), we see that the sensitivity is higher around $\delta_{13} = 90^\circ$ ($-90^\circ$). We have checked that this happens due to the marginalization of $\theta_{24}$ and that the asymmetry disappears if $\theta_{24}$ is fixed.

\begin{figure}
\begin{center}
\includegraphics[width=0.7\textwidth]{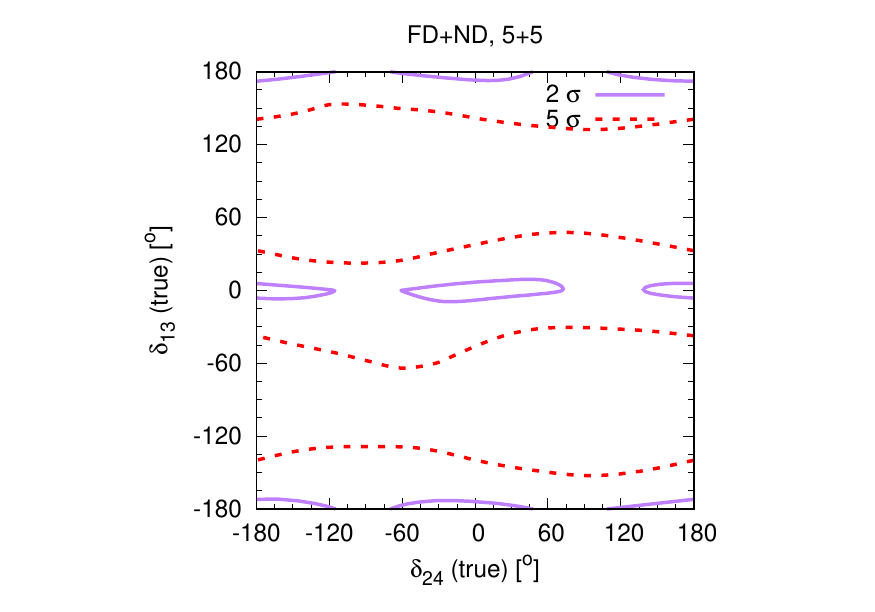} 
\caption{CP violation sensitivity of ESSnuSB in the $\delta_{13}$(true) -- $\delta_{24}$(true) plane. Both $\delta_{13}$ and $\delta_{24}$ are minimized over $0^\circ$ and $180^\circ$ in the test. In addition, $\delta_{34}$ is fixed to $0^\circ$ in both true and test.} 
\label{fig:CP_d13vsd24}
\end{center}
\end{figure}

Next, after investigating the CP violation sensitivity of ESSnuSB due to $\delta_{13}$ only, we study the CP violation sensitivity due to both $\delta_{13}$ and $\delta_{24}$. In Fig.~\ref{fig:CP_d13vsd24}, we show the CP violation sensitivity in the $\delta_{13}$(true) -- $\delta_{24}$(true) plane in terms of 2$\sigma$ and 5$\sigma$ allowed regions, using the FD+ND configuration of ESSnuSB and assuming 5+5. 

Note that in Fig.~\ref{fig:CP_violation}, we marginalized $\delta_{13}$ over $0^\circ$ and $180^\circ$ only and $\delta_{24}$ in the full allowed range, whereas in Fig.~\ref{fig:CP_d13vsd24}, both $\delta_{13}$ and $\delta_{24}$ are minimized over $0^\circ$ and $180^\circ$ in the test. We keep $\theta_{34}$ fixed to $0^\circ$ in both true and test. From Fig.~\ref{fig:CP_d13vsd24}, we see that for $\delta_{13} = \pm 90^\circ$, CP violation can be discovered at $5\sigma$ for all true values of $\delta_{24}$, but for $\delta_{24} = \pm 90^\circ$, CP violation can only be discovered at $5\sigma$ for limited values of $\delta_{13}$. 
This can be understood from Eq.~(\ref{eq:FmeFD}), where we observe that the sensitivity to CP violation comes dominantly from $\delta_{13}$, since the leading-order term is linear in $\mathrm{s} _{13}$, while the sensitivity to $\delta_{24}$ comes from the third-order term in the small parameters $\mathrm{s}_{13}$, $\mathrm{s}_{14}$, and $\mathrm{s}_{24}$. Therefore, the CP violation sensitivity due to $\delta_{13}$ is larger than the CP violation sensitivity due to $\delta_{24}$.
For $\delta_{13} = 0^\circ$ and $\delta_{13} = 180^\circ$, CP violation can be found only at 2$\sigma$ if $\delta_{24} = \pm 90^\circ$. However, for $\delta_{24} = 0^\circ$ and $\delta_{24} = 180^\circ$, CP violation can be observed at more than $2 \sigma$ for all values of $\delta_{13}$, except values around $0^\circ$ and $180^\circ$. Note that in Fig.~\ref{fig:CP_violation}, the CP violation sensitivity for the CP-conserving phases $\delta_{13} = 0^\circ$ and $180^\circ$ is always zero, but in the present case, they can be non-zero, since the sensitivity also comes from $\delta_{24}$.

\section{Summary and conclusions}
\label{sec5}

In this work, we have presented a comprehensive analysis for the sensitivity of the ESSnuSB experiment in the presence of light sterile neutrinos. Assuming both a far (FD) detector and a near (ND) detector, we have analysed the capability of ESSnuSB to constrain the sterile mixing parameters as well as the effect of light sterile neutrinos on the CP violation sensitivity.

In Section~\ref{Sec:ProbLevel}, we have studied the effect of one sterile neutrino flavor at the oscillation probability level, showing that oscillations driven by $\Delta m_{41}^2$ are averaged out at the FD for the values of $\Delta m_{41}^2$ currently preferred~\cite{Dentler:2018sju}. However, oscillations driven by $\Delta m_{41}^2\sim1$~eV$^2$ are fully developed for values of $L/E$ at the ND, allowing to better probe the sterile mixing parameters at the ND, except for the CP-violating phases, whenever the energy reconstruction of the detector is good enough, as shown in Fig.~\ref{fig:ESSnuSB_sens_LSND}. In Section~\ref{Bounds_t14vst24}, we have found that ESSnuSB could be able to probe the interesting parameter space for the LSND result~\cite{Aguilar:2001ty} down to $\sin^2{2\theta_{\mu e}}\sim 10^{-4}$ if there is a perfect energy reconstruction. We have also studied the impact of different shape systematics and the sensitivity could go down to $\sin^2{2\theta_{\mu e}}\sim 10^{-3}$ with the most aggressive choice of systematics. 

Next, we have presented bounds in the $\sin^2\theta_{14}$(test) -- $\sin^2\theta_{24}$(test) plane for different values of $\Delta m_{41}^2$. We have seen that for FD+ND the best bound is obtained for $\Delta m^2_{41} = 1$~eV$^2$ and that the bounds for an ND only are similar to those for FD+ND, implying that much of the sensitivity stems from the ND, as one would expect. In order to estimate the sensitivity with an FD only, we have considered a configuration of FD+ND with no energy information in the ND, so that it behaves as a counting experiment with exactly the same level of systematics as the FD+ND configuration. As expected, at the FD, the strong constraints on the sterile mixing angles for $\Delta m^2_{41} \sim 1$~eV$^2$ are lost and the sensitivity corresponding to different values of $\Delta m^2_{41}$ are similar, given that sterile neutrino driven oscillations are averaged out at the FD. Furthermore, we have found that the sensitivity depends mildly on the neutrino-antineutrino running ratio and the sensitivity is slightly better for the dominant neutrino run. In addition, the bounds on $\theta_{24}$ are slightly better than the bounds on $\theta_{14}$, since there is a stronger dependence on $\theta_{24}$ at the FD.

Then, in Section~\ref{sec:comparison}, comparing the sensitivity of ESSnuSB with that of T2HK, T2HKK, and DUNE, we have found that the sensitivity of ESSnuSB is weaker than the others for most of the parameter space. However, around $\sin^2\theta_{14} = 0.3$ and $\sin^2\theta_{24} < 0.01$, the sensitivity of ESSnuSB is similar to that of DUNE. As the sensitivities of the three others are estimated with an FD only, we have for comparison considered an ESSnuSB configuration with an FD only, having an overall 8~\% (10~\%) systematics in signal (background). This configuration of ESSnuSB gives comparable sensitivity to that of ESSnuSB with FD+ND when studying CP violation, which is the main purpose of the experiment.

Finally, in Section~\ref{sec:CPsensitivity}, we have studied the CP violation sensitivity due to $\delta_{13}$ in presence of light sterile neutrinos and found that if there is an FD only, then the sensitivity to $\delta_{13}$ is greatly reduced compare to the standard case, where there are no sterile neutrinos. Nevertheless, if we consider an ND, then for FD+ND the sensitivity can be even larger than the sensitivity without sterile neutrinos. Note that although the CP violation sensitivity is not coming from the ND itself, its ability to constrain the sterile mixing parameters in turn improves the CP violation sensitivity at the FD. Moreover, we have studied the CP violation sensitivity due to either $\delta_{13}$ or $\delta_{24}$, and found that for any value of $\delta_{24}$ CP violation could be found above $5\sigma$ given that $\delta_{13}\sim \pm 90^{\circ}$, but not the other way around.

The results presented in our work are robust as in calculating the sensitivity and we have scanned the full parameter space consisting of standard and sterile neutrino oscillations. Our results show that in order to estimate the sensitivity to light sterile neutrinos at long-baseline experiments, it is extremely important to include an ND in the experimental setup. 

\section*{Acknowledgements}

We would like to thank Mattias Blennow and Enrique Fernandez-Martinez for useful discussions. We would also like to thank Marie-Laure Schneider and WP6 of the ESSnuSB design study for comments on our work. This work made extensive use of the HPC-Hydra cluster at IFT. This project is supported by the COST Action CA15139 {\it ``Combining forces for a novel European facility for neutrino-antineutrino symmetry-violation discovery''} (EuroNuNet). It has also received funding from the European Union's Horizon 2020 research and innovation programme under grant agreement No 777419. T.O.~acknowledges support by the Swedish Research Council (Vetenskapsr{\aa}det) through Contract No.~2017-03934 and the KTH Royal Institute of Technology for a sabbatical period at the University of Iceland. S.R.~acknowledges support from the ``Spanish Agencia Estatal de Investigaci\'on'' (AEI) and the EU ``Fondo Europeo de Desarrollo Regional'' (FEDER) through the project FPA2016-78645-P and the Spanish MINECO through the Centro de Excelencia Severo Ochoa Program under grant SEV-2016-0597, as well as from the European Union's Horizon 2020 research and innovation programme under the Marie Sklodowska-Curie grant agreements 674896-Elusives and 690575-InvisiblesPlus.

\bibliographystyle{JHEP}
\bibliography{ess_sterile}
  
\end{document}